\documentclass{aa}
\usepackage{psfig,times}
\begin{document}

   \thesaurus{06     
              (02.01.2;  
               08.02.1;  
               08.09.2;  
               08.14.1;  
               13.25.1)}  

   \title{The first radius-expansion X-ray burst from GX\,3+1}

   \author{E.~Kuulkers
           \inst{1,2}
	   \and
           M.~van der Klis
           \inst{3}
          }

   \offprints{Erik Kuulkers}

   \institute{
	      Space Research Organization Netherlands,
	      Sorbonnelaan 2, 3584 CA Utrecht, The Netherlands; E.Kuulkers@sron.nl
              \and
              Astronomical Institute, Utrecht University, 
              P.O.\ Box 80000, 3507 TA Utrecht, The Netherlands
              \and
	      Astronomical Institute ``Anton Pannekoek'', University of
	      Amsterdam, and Center for High Energy Astrophysics, Kruislaan 403,
              1098 SJ Amsterdam, The Netherlands; michiel@astro.uva.nl
             }

   \date{Received --; accepted --}

   \titlerunning{X-ray burst in GX3\,+1}

   \maketitle

   \begin{abstract}

During several observations in 1999 August with RXTE of the low-mass X-ray 
binary GX\,3+1, we found a single short and strong X-ray burst. This is the first
burst from GX\,3+1 which clearly shows evidence for
radius expansion of the neutron-star photosphere during the thermo-nuclear runaway. 
We show that the cooling phase of the neutron star photosphere 
starts already just before the end of the contraction phase.
Considering the fact that the radius expansion is due to 
the burst luminosity being at the Eddington luminosity,
assuming standard burst parameters and accounting for gravitational redshift effects
we derive a distance to the source of $\sim$4.5\,kpc,
although relaxing these assumptions may lead to uncertainties up to $\sim$30\%.
By comparing the persistent flux with that observed at 
the peak of the burst we infer that near the time of the X-ray burst 
the persistent luminosity of GX\,3+1 is $\sim$0.17\,L$_{\rm edd}$, confirming 
predictions from theoretical modeling of X-ray spectra of bright sources
like GX\,3+1.

   \end{abstract}

\section{Introduction}

The overall X-ray intensity of the low-mass X-ray binary (LMXB) GX\,3+1 varies slowly on time scales 
of months to years (Makishima et al.\ 1983; Asai et al.\ 1993, see also Fig.~1).
X-ray bursts in GX\,3+1 were 
discovered by {\em Hakucho}, at a time when the persistent X-ray flux was about half of that 
seen previously (Makishima et al.\ 1983). During that time roughly one burst per day was observed.
The bursts from this source were shown to be thermonuclear flashes
on the neutron star surface, i.e.\ being of type I (Makishima et al.\ 1983;
Asai et al.\ 1993: {\em Ginga}; Molkov et al.\ 1999: {\em GRANAT/ART-P}), 
but none of them showed evidence for photospheric radius expansion. 

GX\,3+1 is one of the four brightest so-called ``atoll'' sources (Hasinger \&\ van der Klis 1989).
The sources in this group (including GX\,13+1, GX\,9+1 and GX\,9+9) hardly 
show any X-ray bursts (if at all),
and display properties like those of other atoll sources when these are in 
their high accretion rate state: their tracks in X-ray colour-colour diagrams
are long, diagonal and slightly curved, while their fast timing properties
are at all times dominated by a relatively weak (1--4\%\ rms) power-law shaped noise component.
Detailed X-ray spectral modeling seems to suggest that they accrete with rates near 
10\%\ of the Eddington mass accretion rate, i.e.\ intermediate between
the more frequently bursting atoll sources and that of the so-called ``Z'' sources
(Psaltis \&\ Lamb 1998). At low accretion rates (and therefore probably low intensities)
such sources are predicted to display the properties characteristic of the more 
frequently bursting atoll sources, which in view of the Hakucho result (see above) 
at least GX\,3+1 seems to satisfy. 

During one of our series of target of opportunity observations with RXTE aimed at observing 
GX\,3+1 at low intensities, we observed a strong ($\sim$2.3\,Crab [2--10\,keV] at maximum) and 
short (15--20\,s) X-ray burst.
The burst onset occurred on 1999 August 10, 18:35:53.5~UTC. In this paper we discuss its
properties.

\section{Observations and Analysis}

Data were acquired with the Proportional Counter Array (PCA; Bradt et al.\ 1993) in various 
observation modes.
During our observation from 1999 August 10, 17:15 to August 11, 00:00~UTC, 
only three units were active, i.e.\ PCU0, PCU2 and 
PCU3. For the spectral analysis of the persistent emission we used
data collected in 16\,s intervals with 129 spectral channels. We accumulated data 
stretches of 96\,s just before and after the burst, combining the three PCU's. 
In order to study the burst properties we used a mode which provides 64 spectral 
channels at a time resolution of 16\,$\mu$s; this mode combines information from all PCU's. 
Spectra during the burst were determined every 0.25\,sec during the first 10\,s, and
every 0.5\,s for the remainder.
All spectra were corrected for background and dead-time using the procedures 
supplied by the RXTE Guest Observer Facility. 
A systematic uncertainty of 1\%\ was taken into account.
For our spectral fits we confined the energy range to 
2.9--20\,keV. The hydrogen column density, N$_{\rm H}$, towards GX\,3+1 was fixed
to that found by the {\em Einstein} SSS and MPC measurements (1.7$\cdot$10$^{22}$\,atoms\,cm$^{-2}$,
Christian \&\ Swank 1997).

Large amplitude, high coherence brightness oscillations have been observed
during various X-ray bursts in other LMXBs (see
Strohmayer 1998, 2000). In our search for possible burst oscillations we
made fast Fourier transforms (FFTs) of data segments ranging from 0.25\,s to 2\,s long
during the burst, with time steps of 0.125\,s (so-called sliding FFTs). We
used the 64 spectral channel and 16 $\mu$sec data set, and limited our
search to the 50 to 2048\,Hz frequency range.  We performed the search in
the whole PCA energy range (2--60\,keV), as well as in a relatively high
energy range (8--60\,keV).

\section{Results}

\subsection{Temporal behaviour}

\begin{figure}
\psfig{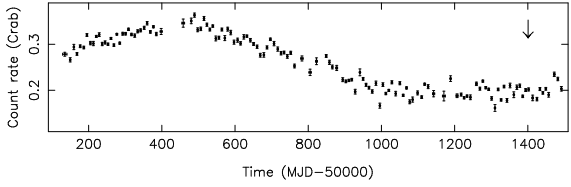}
\caption{ASM light curve (2--12\,keV) of GX\,3+1 from 1996 January 12 to 1999 December 31
in units of the Crab count rate ($\sim$75\,cts\,s$^{-1}$\,SSC$^{-1}$). 
The data points represent the mean of 8 consecutive daily averages. 
The time of the burst observed with the PCA is indicated by an arrow.}
\end{figure}

The light curve of the burst at low energies is single-peaked, whereas at high energies 
it is double-peaked (Fig.~2a,b). The corresponding hardness curve (Fig.~2c)
shows that the burst first hardens, softens, hardens again, and then gradually softens 
again. 
Our search for burst oscillations yielded negative results. Previous burst
oscillations were found mainly during the rise to burst maximum and after
the radius-expansion phase (see Strohmayer 1998, 2000). Using the
0.25\,s long FFTs in the full energy range, we derive upper
limits on the modulation amplitude of burst oscillations of
$\sim$45\%\ at the start of the rise, $\sim$15\%\ at the maximum of the
burst, and $\sim$20\%\ just after the radius expansion phase.
The upper limits in the
8--60\,keV energy band are $\sim$60\%, $\sim$20\%, and $\sim$25\%, respectively.
The 2\,s long FFTs give more stringent upper limits of $\sim$16\%, $\sim$6\%, and
$\sim$9\%, respectively, for the full energy range, whereas we derive
$\sim$24\%, $\sim$7\%, and $\sim$10\%, respectively, in the 8--60\,keV energy band.

\begin{figure}
\psfig{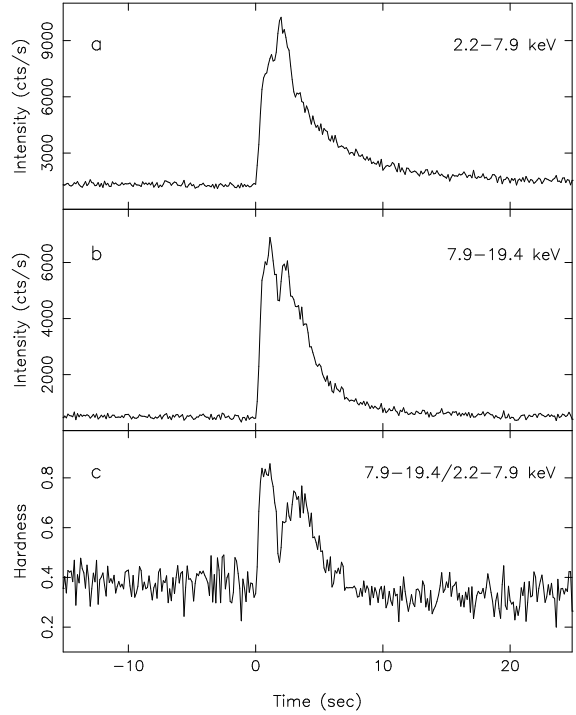}
\caption{The X-ray burst light curve at low (a.) and high (b.) energies and the
corresponding hardness curve (c.), all at a time resolution of 0.125\,sec. 
T=0\,s corresponds to 1999 August 10, 18:35:53.5~UTC.}
\end{figure}

\subsection{Spectral behaviour}

The net burst spectra (i.e.\ total burst spectrum minus 
persistent spectrum) were satisfactorily ($\chi^2_{\rm red}$ less than $\sim$2) 
modeled by black-body emission. 
The results are shown in Fig.~3.
A dip in the black-body temperature, T$_{\rm bb}$, $\sim$2\,s after the 
burst onset is apparent, simultaneous with 
an increase of a factor of $\sim$2 in the black-body radius, R$_{\rm bb}$. 
The total increase/decrease phase of R$_{\rm bb}$ lasts only $\sim$1.5\,sec.

\begin{figure}
\psfig{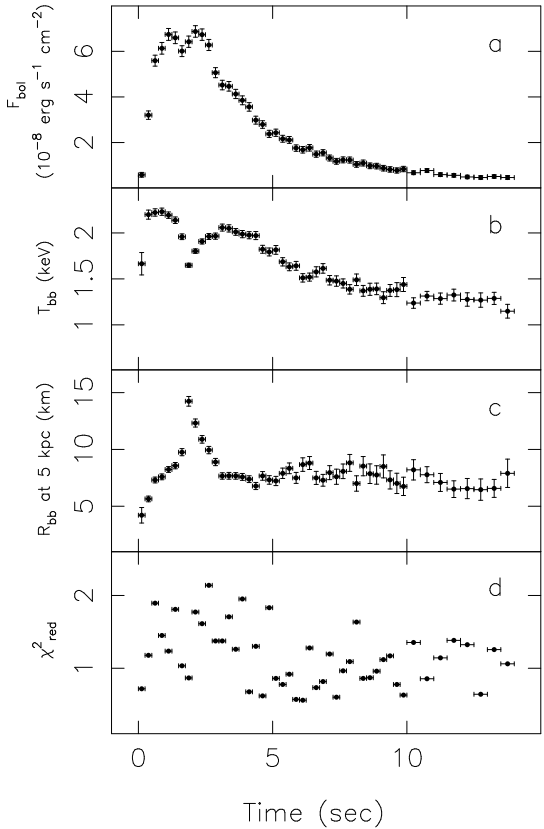}
\caption{Spectral fit results during the burst:
(a) bolometric black-body flux, F$_{\rm bol}$, (b)
black-body temperature, T$_{\rm bb}$, (c) effective black-body radius, R$_{\rm bb}$, at 5\,kpc, 
and (d) goodness of fit expressed in reduced $\chi^2$.}
\end{figure}

We note that the X-ray spectral analysis during bursts can be significantly 
hampered when the persistent emission contains a black-body contribution
from the same surface of the neutron star that emits the burst emission
(van Paradijs \&\ Lewin 1985). In that case our spectral fits to the 
net burst spectra may contain systematic errors 
in the black-body temperature and radius, especially near the end of the burst.
We therefore repeated our spectral analysis to the total burst spectra, fixing the 
non black-body component in the persistent emission
(see Table 1). The black-body component, which now includes all emission 
from the neutron star surface, is left free.
This procedure only slightly alters our estimated black-body flux.
The absence of significant differences between the two methods is mainly due the fact that
the burst is sufficiently stronger than the persistent emission 
(which is reflected by the burst parameter $\gamma$, see below), as also 
noted by Asai et al.\ (1993).

The persistent emission just before and after the burst can be satisfactorily modeled by 
a black-body plus a cut-off power-law component (Table 1). 
Using the X-ray spectral fits
we can determine the peak flux (i.e.\ including persistent emission), 
F$_{\rm peak}$, and the total burst fluence
(i.e.\ the integrated net burst flux), E$_{\rm b}$, and hence the burst parameters
$\gamma$ (=F$_{\rm pers}$/[F$_{\rm peak}$-F$_{\rm pers}$])
and $\tau$ (=E$_{\rm b}$/F$_{\rm peak}$). For the burst parameter 
$\alpha$ (=F$_{\rm pers}$/(E$_{\rm b}$/$\Delta$t)), where $\Delta$t is the time since the
previous burst) we can only give a lower limit, 
since the source is not observed during South Atlantic Anomaly passages and earth occultations. 
However, for a crude estimate we also used 
the mean burst rate of $\sim$2 per day, as observed during the 1999 August to
October BeppoSAX Wide Field Camera campaign (Muller et al.\ 2000, in preparation). 
All burst parameters are also shown in Table~1.

\begin{table}[t]
\caption{Persistent emission and burst properties}
\begin{tabular}{cccl}
\hline\\[-1.5ex]
\multicolumn{4}{l}{Persistent emission$^a$} \\[1.ex]
 & before burst & after burst & unit \\
N$_{\rm H}{}^b$& 
1.7 & 1.7 & 
\multicolumn{1}{l}{$10^{22}$~atoms\,cm$^{-2}$} \\
F$_{\rm pers}$$^c$ & $1.2 \pm 0.2$ & $1.2 \pm 0.2$ & $10^{-8}\,$erg\,s$^{-1}$\,cm$^{-2}$ \\
T$_{\rm bb}$ & $1.34 \pm 0.05$ & $1.38 \pm 0.05$ & keV \\
R$_{\rm bb}$$^d$ & $4.5 \pm 0.3$ & $4.3 \pm 0.3$ & km \\
$\Gamma$ & $1.2 \pm 0.3$ & $1.2 \pm 0.3$ & \\
E$_{\rm cut}$ & $4.9 \pm 0.9$ & $4.7 \pm 1.0$ & keV \\ 
Norm.$^e$ & $1.5 \pm 0.3$ & $1.5 \pm 0.3$ & \\
$\chi^2_{\rm red}$/dof & 1.62/35 & 0.83/35 & \\[1.2ex]
\multicolumn{4}{l}{Burst parameters}\\[1.ex]
F$_{\rm peak}$& 
\multicolumn{2}{c}{$8.1 \pm 0.3$} & 
\multicolumn{1}{l}{$10^{-8}$\,erg\,s$^{-1}$\,cm$^{-2}$} \\
E$_{\rm b}$& 
\multicolumn{2}{c}{$3.52 \pm 0.04$} & 
\multicolumn{1}{l}{$10^{-7}$\,erg\,cm$^{-2}$} \\
$\alpha$& 
\multicolumn{2}{c}{$>6$, $\sim$1360$^f$} &
\multicolumn{1}{l}{} \\
$\tau$& 
\multicolumn{2}{c}{$4.4 \pm 0.2$} & 
\multicolumn{1}{l}{s} \\
$\gamma$& 
\multicolumn{2}{c}{$0.17 \pm 0.03$} & 
\multicolumn{1}{l}{} \\[1.ex]
\hline
\end{tabular}
$^a$\,Absorbed black-body plus cut-off power-law model.\\
$^b$\,Parameter fixed, see text.\\
$^c$\,Unabsorbed persistent flux estimated between 0.01--100\,keV.\\
$^d$\,Effective black-body radius at 5\,kpc.\\
$^e$\,Cut-off power-law normalization (photons~keV$^{-1}$\,cm$^{-2}$\,s$^{-1}$\\ $^{~~}$\,at 1\,keV).\\
$^f$\,Assuming a burst rate of $\sim$2 per day (Muller et al.\ 2000, in prep).
\end{table}

\section{Discussion}

The light curve and X-ray spectral behaviour of the X-ray burst in GX\,3+1 observed with RXTE
show clear evidence for radius expansion of the neutron star photosphere due to near-Eddington 
luminosities during a themonuclear runaway on the neutron star surface
(for a review see e.g.\ Lewin et al.\ 1993). The total time for 
the expansion and contraction phase is only $\sim$1.5\,s, during which the radius
expanded only by a factor of $\sim$2. 
Such short bursts with small expansion phases have been seen in other bright X-ray sources, such
as Cyg\,X-2 (Smale 1998).
The gradual softening at the end of the burst 
is attributed to cooling of the neutron star surface, which is characteristic for
type-I bursts (Hoffman et al.\ 1978). 

During the burst our derived black-body temperatures are smaller, 
whereas our inferred black-body radii (all at the same assumed distance) 
are larger, than reported 
for previous GX\,3+1 bursts
(Makishima et al.\ 1983, see also Inoue et al.\ 1981; Asai et al.\ 1993;
Molkov et al.\ 1999). 
The burst parameter values for $\gamma$ 
($\sim$0.10--0.20) and $\tau$ (4--8\,s) quoted by
Asai et al.\ (1993), and inferred from the observations presented by Makishima et al.\ (1983)
and Molkov et al.\ (1999), are similar to our findings.
We note (see also Asai et al.\ 1993) that $\gamma$, $\tau$ and our estimate of $\alpha$
fall on the extreme end of relations between $\tau$ vs.\ $\gamma$ and $\alpha$ vs.\ $\gamma$
as presented by Van Paradijs et al.\ (1988) for typical type~I bursters.
This shows that if bright sources burst, the burst duration tends to be short 
(order of 10\,s; note however, that some bursts in the bright ``Z'' source 
GX\,17+2 have a duration of the order of minutes, see e.g.\ Kuulkers et al.\ 1997 and references
therein).
It is interesting to note that our X-ray burst from GX\,3+1 is very similar to the radius expansion 
burst seen in Cyg\,X-2 with RXTE in most of its facets, 
except notably for the $\gamma$ 
being a factor 4.3 larger for Cyg\,X-2 (Smale 1998).
Note also that during the burst from Cyg\,X-2 no evidence for pulsations was reported, 
similar to what we conclude for GX\,3+1, both with upper limits on the modulation 
strength which are significantly lower than for bursts during which oscillations were seen
(see Strohmayer 1998, 2000).

\begin{figure}
\psfig{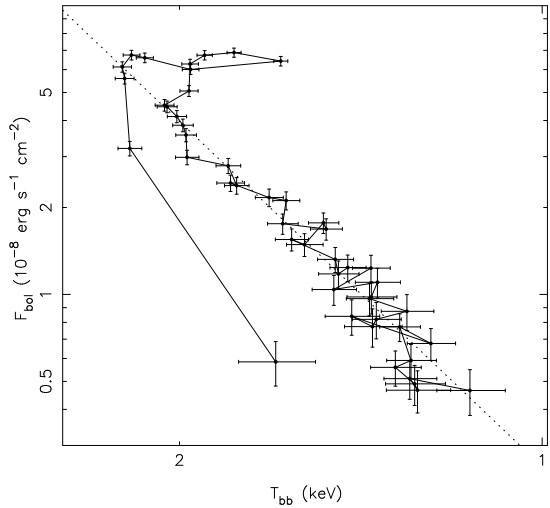}
\caption{Bolometric black-body flux (F$_{\rm bol}$) versus black-body temperature 
(T$_{\rm bb}$) for the first 14\,s of the
burst. Data points are connected for clarity. The dotted line represents the fit to the 
cooling track of the burst, see text. Note that T$_{\rm bb}$ runs from right to left.}
\end{figure}

A convenient way to display the burst properties as they vary in time, is a
flux-temperature diagram, see Fig.~4. In such a diagram 
the phases of expansion/contraction and subsequent cooling of the neutron star photosphere
are distinguished by two separate tracks (see e.g.\ Lewin et al.\ 1993).
GX\,3+1 moves from the middle bottom to 
top left (rising phase), top middle (expansion phase), back towards top left
(contraction phase), and finally to the lower right part of the diagram (cooling phase).
We can adequately fit ($\chi^2_{\rm red}$/dof = 0.9/34) $\log{{\rm F}_{\rm bol}}$ versus
$\log{{\rm T}_{\rm bb}}$ during the cooling phase of the burst by
a straight line with a slope of 3.97$\pm$0.15 (dotted line in Fig.~4).
This means that F$_{\rm bol}$ is consistent with being proportional to 
T$_{\rm bb}^4$, which indicates that the neutron star photosphere
radiates as a black-body during the cooling phase, at a constant radius R$_{\rm bb}$. 
We note that burst spectra are generally not described 
by pure black-body radiation, especially near the Eddington limit 
(see Lewin et al.\ 1993, and references therein). 
Instead the black-body radiation is modified mainly at energies below
$\sim$3\,keV and above $\sim$10\,keV. 
Since our burst spectra are analysed in the 2.9--20\,keV energy range,
we are, therefore, not greatly affected by modified black-body radiation.
Note that we then probably underestimate our bolometric fluxes.

At the start of the expansion phase the black-body bolometric flux and temperature 
values do not match those at the end of the contraction phase.
Our estimated emission areas are the same at these instants; the above then means that the
photosphere is cooler at the end of the contraction phase with respect to the 
start of the expansion phase. From Fig.~4 we see that F$_{\rm bol}$ drops below
the constant peak flux before the end of the contraction phase. We infer that the
cooling phase therefore already started before the end of the contraction phase.

Using the fact that during the expansion and contraction phase of the neutron star
photosphere the burst luminosity equals the Eddington luminosity 
one can get an estimate of the distance (see e.g.\ Lewin et al.\ 1993). 
Assuming standard burst paramaters (isotropy, cosmic abundances and a
canonical neutron star mass of 1.4\,M$_{\sun}$) and taking into account 
gravitational redshift
effects we find $d=4.5\pm 0.1$\,kpc. If we assume a neutron star mass of
2.0\,M$_{\sun}$ we instead find $d=5.1\pm 0.1$\,kpc.
For bright sources like GX\,3+1 most of the hydrogen content is being
burned persistently, so during the expansion/contraction phase the neutron star atmosphere 
is likely to lack hydrogen. Using the Eddington luminosity appropriate for hydrogen-poor
matter then leads to $d=6.1\pm 0.1$\,kpc. Dropping only our assumption 
that the burst radiates isotropically, and assuming anisotropy values of $0.5< \xi <2$
(e.g.\ van Paradijs \&\ Lewin 1987), we derive distances between
3--7\,kpc. On the other hand, if the peak luminosities during radius expansion bursts are standard
candles we can use the mean peak luminosity for such bursts seen in globular cluster sources
for which the distances are known, i.e.\ $3.0 \times 10^{38}$\,erg\,s$^{-1}$ (Lewin et al.\ 1993). 
In this case we derive $d\sim 5.6$\,kpc. These distance estimates 
show that in principle one can get an idea
of the distance to the source, but the exact value still remains rather uncertain
by $\sim$30\%.

The persistent flux during our observations is the same within a factor of $\sim$2 with respect
to the previous reports when GX\,3+1 was bursting, i.e.\ low ($\sim$0.2\,Crab). 
Using the fact that during the peak of the burst the observed (net-burst) luminosity 
is at near Eddington values
we can now for the first time estimate the persistent flux in terms of the Eddington luminosity
for the bright atoll sources like GX\,3+1 (i.e.\ GX\,13+1, GX\,9+1 and GX\,9+9).
For GX\,3+1 we find L$_{\rm pers}\simeq 0.17$\,L$_{\rm edd}$
(assuming the burst and persistent emission is radiated in the same directions). 
This is consistent with that inferred from 
models of X-ray spectra, i.e.\ $\sim$0.1\,L$_{\rm Edd}$ 
(Psaltis \&\ Lamb 1998).

GX\,13+1 has been seen to burst sporadically (Matsuba et al.\ 1995), whereas no bursts have been
reported for GX\,9+1 and GX\,9+9. This may mean that 
GX\,3+1 and GX\,13+1 are accreting near to the critical mass accretion rate at which bursts cease to
occur, whereas GX\,9+9 and GX\,9+1 accrete above this limit. However, this does not explain
the fact that some sources that are accreting at even higher rates (near Eddington), 
i.e.\ Cyg\,X-2 and GX\,17+2, also irreglularly show bursts.

\begin{acknowledgements}
We thank Lars Bildsten, Mariano M\'endez and Dimitrios Psaltis for discussions.
We acknowledge the use of daily averaged quick-look results provided by the ASM/RXTE team.
This work was supported in part by the Netherlands Organization for Scientific Research 
(NWO) grant 614-51-002.
\end{acknowledgements}

\end{document}